%% file: survey.tex
\documentclass[a4paper,UKenglish,cleveref, autoref, thm-restate]{lipics-v2021}
\title{What is Formal Verification without Specifications? A Survey on mining LTL Specifications}

\titlerunning{A Survey on mining LTL Specifications}

\nolinenumbers

\author{Daniel Neider}{TU Dortmund University, Dortmund, Germany, Center for Trustworthy Data Science and Security, University Alliance Ruhr, Dortmund, Germany}{}{}{}

\author{Rajarshi Roy}{Department of Computer Science, University of Oxford, Oxford, UK}{}{}{}

\authorrunning{D. Neider, R. Roy}

\Copyright{Daniel Neider and Rajarshi Roy}

\ccsdesc[500]{Theory of computation}

\keywords{Temporal logic, Passive learning} 

\category{} 

\relatedversion{Full version for published article~\cite{jpk2025}} 

\input{macros}

\begin{document}

\maketitle

\begin{abstract}
	Virtually all verification techniques using formal methods rely on the availability of a formal specification, which describes the design requirements precisely.
	However, formulating specifications remains a manual task that is notoriously challenging and error-prone.
	To address this bottleneck in formal verification, recent research has thus focussed on automatically generating specifications for formal verification from examples of (desired and undesired) system behavior.
	In this survey, we list and compare recent advances in mining specifications in Linear Temporal Logic (LTL), the de facto standard specification language for reactive systems.
	Several approaches have been designed for learning LTL formulas, which address different aspects and settings of specification design.
	Moreover, the approaches rely on a diverse range of techniques such as constraint solving, neural network training, enumerative search, etc.
	We survey the current state-of-the-art techniques and compare them for the convenience of the formal methods practitioners.
\end{abstract}

\input{intro}

\input{prelims}

\input{constraint-based}

\input{enumerative-search}

\input{neural-based}

\input{discussion}

\input{conclusion}

\bibliography{refs}

\end{document}

%% file: macros.tex
\usepackage{lmodern}
\usepackage[T1]{fontenc}
\usepackage{graphicx}
\usepackage{amsmath}
\usepackage{centernot}
\usepackage{amssymb}

\usepackage{graphicx}
\usepackage{mathtools}
\usepackage{tikz}
\usepackage{algorithm}
\usepackage[noend]{algorithmic}
\usepackage{subcaption}
\usepackage{complexity}
\usepackage{wrapfig}
\usepackage{booktabs}
\usepackage{rotating}
\usepackage{array}
\usepackage{ragged2e}
\newcolumntype{P}[1]{>{\RaggedRight\hspace{0pt}}p{#1}}

\newcommand{\LTL}{LTL}

\newcommand{\prop}{\mathcal{P}}

\DeclareMathOperator{\lX}{\mathbf{X}}
\DeclareMathOperator{\lF}{\mathbf{F}}
\DeclareMathOperator{\lG}{\mathbf{G}}
\DeclareMathOperator{\lU}{\mathbf{U}}
\DeclareMathOperator{\lP}{\mathbf{P}}
\DeclareMathOperator{\lS}{\mathbf{S}}
\DeclareMathOperator{\lW}{\mathbf{W}}
\DeclareMathOperator{\lR}{\mathbf{R}}

\newcommand{\lfalse}{\mathit{false}}
\newcommand{\ltrue}{\mathit{true}}

\newcommand{\formula}{\varphi}

\newcommand{\sample}{\mathcal{S}}

%% file: intro.tex

\section{Introduction}

Formal methods refer to the discipline of computer science that employs mathematically rigorous techniques to ensure the safe behavior of software, hardware, and cyber-physical systems.
There have been countless success stories of formal methods, ranging over several application domains such as communication systems~\cite{FeckoUASDMM00,Lowe96}, railway transportation~\cite{BacheriniFTZ06,BadeauA05},
aerospace~\cite{GarioCMTR16,CoferM14}, and operating systems~\cite{VerhulstJ07,KleinAEHCDEEKNSTW10}, to name but a few.
We refer the reader to the exceptional textbook by Baier and Katoen~\cite{model-checking-book} for a comprehensive introduction.

However, there is an important catch with verification techniques: they assume the availability of functional and usable specifications that precisely describe the design requirements.
This assumption is often unrealistic as designing specifications, which had been primarily a manual task, proves not only to be tedious but also error-prone.
Consequently, the availability of formal specifications is widely regarded as a major bottleneck in formal methods~\cite{AmmonsBL02,BjornerH14,Rozier16}.

To overcome this limitation, recent efforts have focused on developing methods that can automatically generate specifications from examples of desired and undesired system behavior.
Notably, a significant body of research has emerged that concentrates on \emph{learning} specifications in Linear Temporal Logic (LTL).
This focus on LTL is due to its dual benefits: mathematical precision and interpretability.
The latter has recently become of increasing interest as it facilitates the application of LTL beyond formal verification to areas such as reinforcement learning, planning, and other AI-related domains~\cite{DBLP:conf/iros/LiVB17,DBLP:conf/ijcai/CamachoIKVM19,DBLP:conf/cdc/HasanbeigKAKPL19,DBLP:conf/icra/Bozkurt0ZP20}.

This survey provides a comprehensive overview of the diverse body of work focused on learning LTL specifications.
In the past decade, researchers have tackled this task from multiple perspectives, spanning various settings and methodologies. As summarized in Table~\ref{tab:related_works}, these efforts can be differentiated based on their learning setup, methodology, and the guarantees they provide, offering a nuanced understanding of the field.

We specifically focus on approaches addressing the \emph{passive learning problem}, where the objective is to learn concise LTL formulas that accurately capture user-provided examples of a system's behavior.
These examples typically consist of two categories: positive (desirable) and negative (undesirable) system behaviors. However, in many practical scenarios, it is unrealistic to assume the availability of perfectly labeled examples for both classes.
Real-world data is often noisy, making it challenging to accurately classify the data.
Furthermore, in many safety-critical domains, such as autonomous vehicles and medical devices, obtaining negative examples may be infeasible or even risky (potentially harming humans).
As a result, some approaches have explored less conventional settings, including noisy data and scenarios where only positive examples are available.

The approaches summarized in Table~\ref{tab:related_works} share a common thread: they employ search strategies to navigate the space of possible LTL formulas, although these strategies differ substantially in their methodology.
Some approaches leverage off-the-shelf constraint solvers (e.g., SAT and SMT solvers) by carefully encoding the learning problem in propositional or first-order logic, while others employ specialized enumeration techniques tailored to the task.
A third category of approaches harnesses advances in deep learning to identify promising LTL formulas.
As a result, the search strategies, considered LTL fragments, and theoretical guarantees vary substantially.

The remainder of this paper compares and contrasts the underlying principles of the aforementioned approaches, categorized into three distinct groups based on their search strategies.
These categories comprise constraint-based, enumeration-based, and neural-network-based methods, which will be explored in Sections~\ref{sec:constraint-based},~\ref{sec:enumeration-based}, and~\ref{sec:neural-based}, respectively.
Section~\ref{sec:preliminaries} provides the required background on LTL and the learning problem we consider.

\begin{sidewaystable}
    \centering
    \caption{Comparison of Related Works and Their Features. `Pos' and `Neg' refer to positive and negative examples, respectively; `SAT', `MaxSAT', `MILP' refer to satisfiability, maximum satisfiability, and mixed-integer linear programming, respectively.}
    \label{tab:related_works}
    \renewcommand{\arraystretch}{1.5}
    \begin{tabular}{*{2}{P{4cm}}*{3}{P{2cm}}p{4cm}}
        \toprule
        \textbf{Work} & \textbf{Primary techniques} & \textbf{LTL Fragments} & \textbf{Classification} &\textbf{Input Data} & \textbf{Guarantees} \\
        \midrule
 Neider and Gavran~\cite{flie}, Riener~\cite{Riener19} & SAT & full LTL & perfect & Pos+Neg & sound, complete, minimal \\
        \midrule
 Camacho and Mcllraith~\cite{CamachoM19} & SAT & full LTL & perfect & Pos+Neg & sound, complete, minimal \\
        \midrule
 Raha et al.~\cite{scarlet} & enumerative search, dynamic programming & comb.\ of directed LTL & perfect/noisy & Pos+Neg & sound\\
        \midrule
 Arif et al.~\cite{ArifLERCT20} & Syntax-Guided Synthesis & past-time LTL & perfect & Pos+Neg & sound, complete, minimal\\
        \midrule
 Gaglione et al.~\cite{GaglioneNRTX21} & MaxSAT & full LTL & noisy & Pos+Neg & sound, complete, minimal \\\hline
 Chou et al.~\cite{ChouOB22} & MILP, counterexample-guided & full LTL & perfect & Pos only & sound, complete \\
        \midrule
 Ghiorzi et al.~\cite{GhiorziCPBTN23} & enumerative search & full LTL & perfect & Pos+Neg & sound, complete, minimal\\
        \midrule
 Ielo et al.~\cite{IeloLFRGR23} & Answer Set Programming & full LTL & perfect & Pos+Neg & sound, complete, minimal\\
        \midrule
 Roy et al.~\cite{ltl-from-positive-only} & SAT, counterexample-guided & full LTL & perfect & Pos only & sound, complete, language minimal\\
        \midrule                
 Valizadeh et al.~\cite{LTL-GPU} & enumerative search, GPU acceleration & full LTL & perfect/noisy & Pos+Neg & sound \\
        \midrule
 Luo et al.~\cite{DBLP:conf/aaai/LuoLDWPZ22} & Graph Neural Network & full LTL & noisy & Pos+Neg & No \\
        \midrule
 Wan et al.~\cite{DBLP:conf/aaai/WanLDLYP24} & Graph Neural Network & full LTL & noisy & Pos+Neg & No \\
        \bottomrule
    \end{tabular}

\end{sidewaystable}

%% file: prelims.tex

\section{Preliminaries}
\label{sec:preliminaries}

\subsection{System Executions and Words}
Informal methods, executions or trajectories of systems are typically formalized as sequences of symbols from a finite non-empty set $\Sigma$, known as alphabet.
We refer to such sequences as \emph{words} over $\Sigma$.
A word $w= a_0 a_1 \dots$, where $a_i\in \Sigma$, can be either finite or infinite, depending on whether the execution it represents is finite or infinite.
The set of infinite words over $\Sigma$ is denoted by $\Sigma^\omega$, while the set of finite words is denoted by $\Sigma^*$.
Given a word $w=  a_1 a_2 \dots$ in $\Sigma^\ast$ or $\Sigma^\omega$, we let $w[i] \coloneqq a_i$ denote symbol of $w$ at position~$i$ and $w[i:] \coloneqq a_i a_{i+1} \dots $ the suffix starting from the starting from position~$i$.
The length $|w|$ of a word $w$ is the number of its symbols.
In particular, the empty word, denoted by $\varepsilon$, has length zero.

\subsection{Linear Temporal Logic (LTL)}
The logic \LTL{}~\cite{DBLP:conf/focs/Pnueli77} is the de~facto standard for reasoning about executions, or sequences of events, of reactive systems.
Typically, specific events in a system are abstracted using a set $\prop$ of propositions, which represent events or properties of interest in the system under consideration.
A system execution is then modeled by a word over the alphabet $\Sigma = 2^\prop$, capturing the propositions that hold true at specific time points along the system's execution.

Given a set $\prop$ of propositions, the syntax of \LTL{} formulas is defined inductively using the grammar
$$\formula \coloneqq p\in\prop \mid \lnot\formula \mid \formula \lor \formula \mid \lX\formula \mid \formula \lU \formula,$$
where $\lnot$ (not) and $\lor$ (or) are Boolean operators, while $\lX$ (neXt) and $\lU$ (Until) are so-called future-time temporal operators.
Several derived Boolean operators, including $\land$ (conjunction) and $\rightarrow$ (implication), as well as temporal operators, such as $\lF$ (Eventually), $\lG$ (Always), $\lW$ (Weak Until), and $\lR$ (Release), are often added as syntactic sugar.
Additionally, some works also incorporate past-time temporal operators, including $\lP$ (Previously) and $\lS$ (Since), which are past-time analogs of $\lX$ and $\lU$, respectively.

To define the semantics of LTL, one usually uses a ``model relation'', denoted by $\models$, which captures when the suffix of a word $w \in (2^\prop)^\omega$ starting at a position $i\in \mathbb N$ satisfies an LTL formula $\varphi$.
Formally, this relation is given as follows:
\begin{align*}
    \hspace{3.5cm} w,i\models p & \text{  if and only if  } p\in w[i] \\
 w,i\models \neg\varphi & \text{  if and only if  } w,i\not\models \varphi \\
 w,i\models \varphi_1\lor\varphi_2 & \text{  if and only if  } w,i\models \varphi_1 \text{ or } w,i\models \varphi_2 \\
 w,i\models \lX\varphi & \text{  if and only if  } w,i+1\models \varphi\\
 w,i\models \varphi_1\lU\varphi_2 & \text{  if and only if } w,j \models \varphi_2 \text{ for some $i\leq j$ and} \\
 &\hspace{1cm} w,k\models \varphi_1 \text{ for each $i \leq k < j$}
\end{align*}
If the entire word starting at position $0$ satisfies $\varphi$ (i.e., $w, 0 \models \varphi$), we simply write $w\models\varphi$ and say that $w$ satisfies $\varphi$.

Traditionally, LTL has been interpreted over infinite words, but there is a growing interest in interpreting LTL over finite words~\cite{GiacomoV13}, particularly in artificial intelligence applications. To reflect this shift, it is sufficient to adapt the model relation slightly, taking into account the end of a word in the operators $\lX$ and $\lU$.
More precisely, we modify the model relation for a finite word $w \in (2^\prop)^\ast$ as follows:
\begin{align*}
 \hspace{2cm}w,i\models \lX\varphi & \text{  if and only if $i < |w|-1$ and } w,i+1 \models \varphi\\
 w,i\models \varphi_1\lU\varphi_2 & %
        \begin{multlined}[t]
            \text{  if and only if } w,j \models \varphi_2 \text{ for some $ i \leq j \leq |w-1|$ and} \\
 w,k\models \varphi_1 \text{ for each $i \leq k < j$}
        \end{multlined} 
\end{align*}


The size of an LTL formula $\varphi$, denoted by $|\varphi|$, is defined as the number of its unique subformulas.
For example, the size of $\varphi \coloneqq (p \lU \lG q) \lor \lX(\lG q)$ is six since it contains six unique subformulas: $(p \lU \lG q) \lor \lX(\lG q)$, $p \lU \lG q$, $\lX(\lG q)$, $\lG q$, $p$, and $q$.

When learning LTL formulas, it is useful to have a canonical representation.
A common approach is to use a so-called \emph{syntax directed acyclic graph (DAG)}, which is a syntax tree that merges common subformulas.
Figure~\ref{fig:ltl-reps} illustrates the distinction between a syntax tree (Figure~\ref{subfig:syntax-tree}) and a syntax DAG (Figure~\ref{subfig:syntax-dag}) for the formula $\varphi \coloneqq (p \lU \lG q) \lor \lX(\lG q)$.
Notably, the size of a formula and the number of nodes in its syntax DAG coincide, whereas the number of nodes in the syntax tree can be exponentially larger.

\begin{figure}
    \centering
    \subcaptionbox{Syntax Tree\label{subfig:syntax-tree}}{
        \begin{subfigure}[b]{0.4\textwidth}
            \centering
            \begin{tikzpicture}
            \node (1) at (0, 0) {$\lor$};
            \node (2) at (-.6, -.9) {$\lU$};
            \node (3) at (.6, -.9) {$\lX$};
            \node (4) at (-1.2, -1.8) {$p$};
            \node (5) at (-.1, -1.8) {$\lG$};
            \node (6) at (-.1, -2.7) {$q$};
            \node (7) at (.6, -1.8) {$\lG$};
            \node (8) at (.6, -2.7) {$q$};
            \draw[->] (1) -- (2); 
            \draw[->] (1) -- (3);
            \draw[->](2) -- (4);
            \draw[->] (2) -- (5);
            \draw[->] (3) -- (7);
            \draw[->] (5) -- (6);
            \draw[->] (7) -- (8);
            \end{tikzpicture}
        \end{subfigure}
 }
    \subcaptionbox{Syntax DAG\label{subfig:syntax-dag}}{
        \begin{subfigure}[b]{0.4\textwidth}
            \centering
            \begin{tikzpicture}
            \node (1) at (0, 0) {$\lor$};
            \node (2) at (-.6, -.9) {$\lU$};
            \node (3) at (.6, -.9) {$\lX$};
            \node (4) at (-1.2, -1.8) {$p$};
            \node (5) at (0, -1.8) {$\lG$};
            \node (6) at (0, -2.7) {$q$};
            \draw[->] (1) -- (2); 
            \draw[->] (1) -- (3);
            \draw[->](2) -- (4);
            \draw[->] (2) -- (5);
            \draw[->] (3) -- (5);
            \draw[->] (5) -- (6);
            \end{tikzpicture}
        \end{subfigure}
 }
    \caption{Representations of LTL formula $\varphi=(p \lU \lG q) \lor \lX(\lG q)$}
    \label{fig:ltl-reps}
\end{figure}

\section{Passive Learning of LTL Formulas.}
With the necessary groundwork established, we now turn our attention to the central task of this survey: learning LTL formulas from examples.
To this end, we assume that the examples of desired and undesired system executions are bundled in a \emph{sample}, denoted by $\sample$.
In the standard setting for passive learning of LTL formulas, this sample takes the form of a pair $\sample=(P,N)$, comprising two sets of words: $P$, consisting of positive examples, and $N$, comprising negative examples, where $P\cap N=\emptyset$.

A crucial concept in this setting is that of \emph{consistency}, where an LTL formula $\varphi$ is said to be consistent with the sample $\sample$ if it satisfies two conditions: firstly, every word $u\in P$ must satisfy $\varphi$ (i.e., $u\models\varphi$), and secondly, every word $v\in N$ must not satisfy $\varphi$ (i.e., $v\not\models\varphi$).
This definition allows us to define the passive learning task formally.

\begin{definition}[Passive Learning of LTL formulas]
    \label{def:passive-learning}
 Given a sample $\sample=(P,N)$, compute a minimal LTL formula $\varphi$ that is consistent with $\sample$.
\end{definition}

A crucial aspect of the above definition is the minimality requirement for the prospective LTL formula, as emphasized by Neider and Gavran~\cite{flie}.
Notably, the problem becomes trivial if this size restriction is relaxed: for any $u \in P$ and $v \in N$, one can construct a formula $\varphi_{u,v}$ that captures the first symbol where $u$ and $v$ differ using a sequence of $\lX$-operators and a suitable propositional formula, ensuring that $u\models \varphi_{u,v}$ and $v\not\models \varphi_{u,v}$.
Then, the conjunction of these formulas, $\bigwedge_{u\in P} \bigvee_{v\in N} \varphi_{u,v}$, is consistent with the sample $\sample$.
However, a formula that simply enumerates the differences in the positive and negative examples suffers from overfitting and fails to generalize the temporal patterns.
Furthermore, the resulting formula can become excessively complex, thereby compromising its interpretability.

There has been significant research aimed at understanding the theoretical aspects of the above problem, particularly in terms of computational complexity~\cite{DBLP:conf/icgi/FijalkowL21,DBLP:journals/corr/abs-2312-16336,DBLP:journals/corr/abs-2408-04486}. In this survey, however, we will primarily focus on practical algorithms that efficiently address the problem.
While the problem as stated represents the standard passive learning framework, real-world scenarios often deviate from this idealization. 
In practice, samples may be imperfect—either noisy (e.g., due to misclassifications or sensor reading errors) or incomplete (e.g., when only positive examples are available). 
To account for such challenges, numerous studies have extended the passive learning framework to accommodate these scenarios.


%% file: constraint-based.tex

\section{Constraint-Based Approaches}
\label{sec:constraint-based}

As hinted at in the introduction, constraint-based approaches leverage off-the-shelf solvers to search for prospective LTL formulas.
These solvers employ a wide range of technologies, including (i) solvers for satisfiability (SAT) \cite{DBLP:conf/tacas/MouraB08,DBLP:conf/cav/BarrettCDHJKRT11} and maximum satisfiability (MaxSAT) \cite{DBLP:conf/sycss/BjornerP14} for propositional logic, (ii) Inductive Logic Programming (ILP) \cite{DBLP:journals/ngc/GelfondL91}, (iii) Mixed Integer Linear Programming (MILP) \cite{DBLP:books/daglib/0090563}, and (iv) Syntax-Guided Synthesis (SyGuS)~\cite{DBLP:conf/fmcad/AlurBJMRSSSTU13}.

At the heart of constraint-based approaches lies the idea of translating the learning problem into one or several satisfiability problems within a suitable logical framework (e.g., SAT or SMT).
Although the solver technologies may differ, the underlying logical encodings of the learning problem exhibit striking similarities across most works.
In particular, most approaches separate the encoding of the syntax of the prospective formula from its semantics, allowing for a flexible and modular formulation of the search problem.
This modularity enables the customization of the search to accommodate specific requirements, such as targeting a particular subclass of formulas or satisfying a specific subset of examples.

To further illustrate this approach, let us consider one of the pioneering works in this category, the paper by Neider and Gavran~\cite{flie}, which leverages SAT solving.
The core of their approach revolves around constructing a series $(\Phi^{\sample}_n)_{n=1,2,\dots}$ of propositional formulas that facilitate the search for prospective LTL formulas of increasing size $n$.
Specifically, each formula $\Phi^{\sample}_n$ satisfies two crucial properties:
(i) it is satisfiable if and only if there exists an LTL formula of size $n$ that is consistent with $\sample$ and (ii) a satisfying assignment of $\Phi^{\sample}_n$ contains sufficient information to construct such an LTL formula.
By incrementally increasing the value of $n$ until $\Phi^{\sample}_n$ becomes satisfiable, one can obtain an LTL formula that is guaranteed to be minimal and consistent with $\sample$.

The formula $\Phi^{\sample}_n$ is constructed as a conjunction of two subformulas, $\Phi^{\sample}_n = \Phi^{DAG}_n \wedge \Phi^{con}_n$.
The subformula $\Phi^{DAG}_n$ encodes the syntax DAG of the prospective LTL formula and encompasses a range of constraints, ensuring that fundamental properties of a syntax DAG of an LTL formula are satisfied (e.g., each node being labelled by a unique LTL operator and each node having at most two children).
%
%
On the other hand, the subformula $\Phi^{con}_n$ ensures that the the positive examples satisfy the prospective LTL formula while the negative ones violate it. 
To this end, the formula $\Phi^{con}_n$ encodes of the semantics of LTL on the given positive and negative words similar to Bounded Model Checking~\cite{DBLP:journals/fmsd/ClarkeBRZ01}.
They rely on the SAT solver Z3 \cite{DBLP:conf/tacas/MouraB08} for their implementation.

Riener~\cite{Riener19} expanded upon the work of Neider and Gavran, streamlining the SAT encoding by preprocessing the possible syntactic structures of LTL formulas, formalized as \emph{partial DAGs}.
By precomputing and storing partial DAGs, which are not yet labeled with LTL operators, Riener's approach enables a more efficient search by decomposing the search space according to the underlying DAG structure.
This innovation potentially enables parallelization, thereby accelerating the encoding process and reducing computational complexity.

Camacho and Mcllraith~\cite{CamachoM19} propose a SAT encoding similar to that of Neider and Gavran, leveraging Alternating Finite Automata (AFA), a type of finite state acceptors for words.
Since the definition of AFA are distinct from LTL, both approaches seem different at the first glance.
However, it is well established that counter-free AFA, a specific subclass of AFA, are equivalent to LTL in terms of their expressive power.
Furthermore, the syntactic structure and semantic interpretation of counter-free AFA, as demonstrated by Camacho and Mcllraith~\cite[Theorem 1, Property 1]{CamachoM19}, show a striking resemblance to those of LTL, ultimately leading to an encoding that is almost identical to that of Neider and Gavran.
They rely on the SAT solver Pycosat \cite{DBLP:journals/jsat/Biere08} for their implementation.

Arif et al.~\cite{ArifLERCT20} elevate the SAT-based encoding of Neider and Gavran to a syntax-guided synthesis (SyGUS) framework.
Such frameworks, employed in programming synthesis, inherently support various search heuristics, including symmetry breaking, rewrite rules, and others.
Notably, in contrast to previous work that translates LTL semantics to SAT in a straightforward manner~\cite{flie}, Arif et al.\ rely on a bit-vector arithmetic-based encoding.
They use the CVC4SY solver \cite{DBLP:conf/cav/ReynoldsBNBT19} for their implementation.

Ielo et al. \cite{IeloLFRGR23} elevate the SAT-based encoding of Nieder and Gavran to the Answer Set Programming (ASP) framework, a declarative programming paradigm that allows for defining problems in terms of rules and constraints.
They devise two formulations of the passive learning problem in ASP, as an abduction problem and as a context-depedent learning problem.
They rely ASP solvers such as CLINGO and ILASP \cite{DBLP:series/synthesis/2012Gebser} in their implementation.

\subsection{Learning from Noisy Data.}
To accommodate noisy data, a relaxation of the requirement for the generated formula to be consistent with all examples is necessary.
This relaxed consistency criterion is often expressed using metrics of misclassification, such as the loss function $$l(\sample, \varphi) = \frac{\sum_{u\in P} [u\not\models \varphi] + \sum_{v\in N} [v\models\varphi]}{|P|+|N|},$$ where $\sample = (P, N)$ and the Iverson bracket $[\phantom{x}]$ maps true statements to $1$ and false to $0$.
This loss quantifies the proportion of examples misclassified by the formula $\varphi$ and closely mimic standard loss functions used in statistical machine learning.

To learn minimal LTL formulas that minimimize the above loss function, Gaglione et al.~\cite{GaglioneNRTX21} propose translating the problem into a Maximum Satisfiability (MaxSAT) instance, mirroring the techniques employed by Neider and Gavran~\cite{flie} and Riener~\cite{Riener19} for propositional logic.
MaxSAT extends the classical satisfiability problem of propositional logic, allowing for the definition of hard constraints (mandatory clauses) and soft constraints (optional clauses).
The solution to a MaxSAT problem is a variable assignment that satisfies all hard constraints and as many soft constraints as possible.
Gaglione et al.\ capitalize on this technology by designating all clauses in $\Phi^{DAG}_n$ as hard constraints and selected clauses in $\Phi^{con}_n$ as soft constraints.
As a result, they obtain a minimal LTL formula that minimizes the specified loss function.
They rely on the MaxSAT solving capabilities of Z3 \cite{DBLP:conf/sycss/BjornerP14} for their implementation.

In fact, by following a similar method, almost all of the constraint-based approaches can potenitally be extended to noisy settings if the solver employed allows such relaxations.

\subsection{Learning from Positive Examples Only.}
The problem of learning from positive examples only is a special case of the one-class learning task, where only one class of inputs (positive or negative) is available. This problem frequently arises in AI applications, particularly in the context of explainability, where one seeks to infer the behavior of an autonomous agent from observational data.

Unlike learning from noisy data, extending constraint-based approaches to learn from only positive examples is not straightforward.
The primary reason for this is that learning LTL formulas from positive examples is an inherently ill-posed problem.
Given a set of positive examples $P$, the smallest LTL formula that is consistent with $P$ is the trivial formula $\ltrue$, which is satisfied by any word.
Clearly, this formula is too general and does not provide any insights into the underlying (temporal) patterns in the examples.

To address this challenge, Roy et al.~\cite{ltl-from-positive-only} propose strongness---or specificity---as an additional optimization parameter besides the size of the formula.
In particular, the authors solve formulate a learning task wherein, given a set $P$ of positive examples and a size bound $n > 0$, the goal is to learn an LTL formula $\varphi$ that satisfies the following three conditions: (i) each $w \in P$ satisfies $\varphi$, (ii) $\varphi$ has size at most $n$, and (iii) there exists no other formula with the former two properties that implies $\varphi$.

To tackle this problem, Roy et al.\ employ a counterexample-guided inductive synthesis loop~\cite{DBLP:conf/fmcad/AlurBJMRSSSTU13}, which leverages negative examples to guide the learning algorithm towards a most specific LTL formula.
In each iteration of the loop, the authors utilize one of the aforementioned SAT-based methods to construct a consistent LTL formula.
This formula is then analyzed, and if necessary, used to generate a new negative example that directs the search towards a more specific formula.
This iterative process continues until no formulas more specific than the current formula can be found, at which point the algorithm terminates.

Chou et al. \cite{ChouOB22} propose learning LTL formulas from positive examples of high-dimensional data.
They rely on domain specific non-convex cost functions involving the positive examples and LTL formulas to ensure that the prospective formula tightly describes the examples.
To search for the prospective formula, they provide a joint encoding of the cost functions and the syntax DAG of LTL formulas using Mixed Integer Linear Programming (MILP).
They rely on the MILP solver IPOPT \cite{WachterB06} for their implementation.

%% file: enumerative-search.tex

\section{Enumeration-Based Approaches}
\label{sec:enumeration-based}

The constraint-based approaches discussed in Section~\ref{sec:constraint-based} provide a systematic method for learning arbitrary LTL formulas. 
However, the performance of these approaches is limited by the capabilities of the underlying solvers.
The search techniques typically employed in the solvers are not optimized for learning LTL formulas, thus often leading to bottlenecks in scalability.

As a result, recent works have started exploring alternative search strategies that are tailored to navigate through the search space of LTL formulas efficiently.
These approaches search through relevant/interesting LTL formulas in a more targeted manner that results in scalability, typically at the expense of the minimality of the learned formulas.

A prominent example of this approach is Scarlet, a tool developed by Raha et al.~\cite{scarlet,DBLP:journals/jossw/RahaRFN24} that detects and accumulates common temporal patterns in a given sample.
For instance, analyzing a sample consisting of a positive word $u = \{p\}\{p\}\{q\}\{p\}\{r\}\{p\}$ and a negative word $v = \{p\}\{p\}\{r\}\{p\}\{q\}$, Scarlet extracts the formula $\lF(q\wedge\lF(r))$, which captures the order in which the propositions $q$ and $r$ appear.
By employing dynamic programming, the tool identifies a large number of such patterns, which are then translated into a simple yet expressive LTL fragment named directed LTL.
Scarlet then combines a suitable selection of directed LTL formulas to construct a consistent formula using a novel procedure called Boolean subset cover.
Unlike constraint-based approaches, Scarlet's search strategy integrates syntax and semantics computations in a single, unified process, resulting in a more efficient and effective method for learning LTL formulas.

Another notable example is the highly parallelized algorithm developed by Valizadeh et al.~\cite{LTL-GPU}, which is designed to leverage the processing power of Graphics Processing Units (GPUs).
Their approach comprises two pivotal procedures: \emph{relaxed unique checks} (RUCs) and \emph{divide and conquer} (D\&C).
The RUCs procedure performs a bottom-up search through the syntax of LTL formulas, eliminating redundant formulas that exhibit the same behavior on the given sample.
Since this procedure is resource-intensive, it cannot be easily extended to large samples.
To mitigate this, the D\&C procedure partitions the sample into smaller, manageable subsets on which RUCs can be applied in parallel.
The resulting formulas can then be combined using Boolean combinations to generate one consistent LTL formula.
Internally, Valizadeh et al.'s approach employs bit-vectors to encode the semantics of LTL formulas, which can be highly efficiently implemented on GPUs.
By exploiting the parallel processing capabilities of GPUs, the authors achieve a significant speedup, making their approach perhaps the most scalable one of all.

Ghiorzi et al.~\cite{GhiorziCPBTN23} propose a range of heuristics to expedite the enumeration of LTL formulas.
Inspired by Riener~\cite{Riener19}, the authors first employ an enumeration strategy based on partial DAGs to navigate the search space quickly.
Then, they utilize LTL rewrite rules to eliminate equivalent and redundant formulas, leveraging rules such as $\varphi \wedge \neg \varphi \equiv \lfalse$ and $\neg \lF\varphi \equiv \lG(\neg \varphi)$.
Additionally, the authors efficiently eliminate tautologies and contradictions by alternately checking the satisfaction of enumerated LTL formulas on positive and negative examples.

%% file: neural-based.tex

\section{Neural Network-Based Approaches}
\label{sec:neural-based}

Recent research also focuses on leveraging the optimized training capabilities of neural networks to achieve scalability in the LTL learning process. 
However, due to the inherent uncertainty of neural network training, these approaches lack theoretical guarantees regarding the consistency of the learned LTL formulas. 
Nonetheless, they can produce reasonably good LTL formulas from large, typically noisy datasets.

The current approaches specifically exploit Graph Neural Networks (GNNs)~\cite{DBLP:journals/debu/HamiltonYL17} to learn LTL formulas.
GNNs are a powerful neural architecture that learns vector representations of vertex and edge features, typically called embeddings.
More formally, GNNs define a message-passing process between vertices in a graph, where each vertex aggregates information from its neighbors to update its own representation.
This process is repeated several times, allowing the model to learn complex patterns and relationships between vertex and edge features.

The critical insight to understanding the connection of LTL and GNNs is to view a word $w=a_1 a_2 \dots a_n$ as a linear graph $v_{1}\rightarrow v_{2} \dots \rightarrow v_n$ with $n$ nodes.
This representation allows associating a feature vector $x_i$ to each node $v_i$ that tracks the satisfaction of the different subformulas of a prospective LTL formula $\varphi$ when evaluated at the $i$-position of an example (see the definition of the model relation on Page~4).
By leveraging message passing, the satisfaction of the entire formula $\varphi$ can be computed by aggregating the feature vectors of nodes $v_j$ with $j > i$ according to the semantics of LTL.


Luo et al.~\cite{DBLP:conf/aaai/LuoLDWPZ22} built upon the insight of representing words as linear graphs to train a GNN on a sample $\sample = (P, N)$.
Subsequently, the authors use the learned network weights to extract an LTL formula that closely approximates the behavior of the GNN on the given sample. 

Although Luo et al.'s work pioneered the use of GNNs for learning LTL formulas, it suffers from the limitation that the extracted LTL formula might accurately capture the behavior of the trained GNN.
Wan et al.~\cite{DBLP:conf/aaai/WanLDLYP24} address this shortcoming by introducing an enhanced architecture.
In particular, the authors devise a faithful encoding of the LTL semantics within the GNN architecture, achieved through parametric constraints on the network weights.
This innovative encoding ensures that a consistent LTL formula can always be reliably extracted from the trained GNN.

%% file: discussion.tex

\section{Other Settings}

Our discussion thus far has centered around the classical passive learning problem for LTL as defined in Definition~\ref{def:passive-learning}.
However, several variants of this problem have been explored, each presenting unique challenges.
In this section, we discuss three such extensions, highlighting their distinct characteristics and proposed solutions.

\subsection{Mining LTL based on Templates}

A key property of Definition 1 is that it makes no restrictions on the syntactic structure of an LTL formula as long as this formula is consistent with the given sample.
In practice, however, users sometimes want to incorporate domain knowledge into the learning process or must confine the solutions to specific LTL fragments.
Unfortunately, the methods discussed in Sections~\ref{sec:constraint-based} to \ref{sec:neural-based} do not offer this level of fine-grained control and, thus, cannot be applied in such situations.

To address this limitation, researchers have proposed methods that allow users to provide a partial formula, typically referred to as a template or sketch, where parts can be omitted (typically indicated by a question mark).
In this framework, the task of a learning algorithm is then to infer the missing part of a template (or sketch) so that the completed formula is consistent with a given set of examples.
Notable examples operating in this setting are the methods by Lemieux et al.~\cite{DBLP:conf/kbse/LemieuxB15} and Lutz et al.~\cite{LutzNR23}.
The former approach considers templates, called \emph{property types}, in which question marks serve as placeholders for missing propositions.
By contrast, the latter approach introduces so-called \emph{sketches} in which placeholders can represent missing operators and even entire subformulas.

For example, Li et al.~\cite{LiDS11} have developed a method to mine LTL specifications in the GR(1) fragment of LTL, using templates that include $\lG\lF ?$, $\lG(? \rightarrow \lX ?)$, and others.
Similarly, Shah et al.~\cite{ShahKSL18} have focused on conjunctive LTL formulas built from a limited set of typical temporal properties collected by Dwyer et al.~\cite{DBLP:conf/fmsp/DwyerAC98}.
Another notable example is the work by Kim et al.~\cite{KimMSAS19}, which considers a set of interpretable LTL templates originating in software system development and seeks to infer formulas robust to noise in the input data.

\subsection{Mining LTL from Natural Language}

One of the significant barriers to adopting temporal logic in practice is the limited expertise of practitioners and engineers in this area \cite{Holzmann02,GreenmanSNK23}.
As a result, they often prefer to specify their requirements in natural language, which is more intuitive and accessible to them.
Several research efforts have focused on bridging this gap by automatically extracting LTL formulas from natural language descriptions.
Early approaches~\cite{FinucaneJK10,Kress-GazitFP08,NikoraB09,GhoshELLSS16,GiannakopoulouP20a} achieved this by efficiently parsing English sentences to translate them into LTL and other temporal logic formulas.
The advent of data-driven techniques has led to the development of neural-network-based methods~\cite{OhPNHPT19,HahnSKRF21,Cherukuri0S22} that rely on human-labeled pairs of natural language descriptions and corresponding logic formulas.
More recently, researchers have begun to leverage the impressive natural language understanding capabilities of Large Language Models to enhance the translation capabilities further~\cite{CoslerHMST23,PanCB23,lang2LTL,nl2ltl}, offering promise for more effective and efficient property specification.

\subsection{Logics beyond LTL}

The widespread adoption of continuous-time logics, such as Signal Temporal Logic~(STL), in the context of cyber-physical systems has spawned a significant body of research focused on learning specifications in STL.
In fact, a comprehensive survey by Bartocci et al.~\cite{BartocciMNN22} is dedicated entirely to this problem.
Most of these works concentrate on learning formulas with a specific syntactic structure~\cite{dtbombara,dtmethod} or identifying time intervals for given STL formulas~\cite{asarin,rpstl1,rpstl2}.
A handful of works also tackle the more general passive learning problem, where the goal is to learn STL formulas of arbitrary structure~\cite{MohammadinejadD20,genetic}.

In addition to linear-time properties, there exist several works focusing on learning branching-time properties in Computation Tree Logic~(CTL).
For instance, Chan~\cite{Chan00} addresses the problem of completing simple CTL templates, while Wasylkowski and Zeller~\cite{WasylkowskiZ11} investigate inferring operational preconditions for Java methods in CTL.
Recent research by Pommellet et al.~\cite{DBLP:conf/ijcar/PommelletSS24} and Bordais et al.~\cite{DBLP:conf/fm/BordaisNR24} demonstrate that constraint-based techniques can be used to learn not only CTL but also Alternating-time Temporal Logics (ATL), which extends CTL for multi-agent systems.

Apart from LTL, STL and CTL, there is also research on learning formulas in other temporal logics, such as Metric Temporal Logic~(MTL)~\cite{DBLP:conf/vmcai/RahaRFNP24}, and the Property Specification Language~(PSL)~\cite{0002FN20}.

%% file: conclusion.tex

\section{Conclusion}

This survey provides a comprehensive overview of the diverse research efforts focused on learning specifications in temporal logic, with a particular emphasis on Linear Temporal Logic.
We systematically compared and contrasted these works based on their search strategies to navigate the vast space of possible formulas.
Some approaches leverage off-the-shelf solvers, while others propose customized enumeration techniques or exploit advances in deep learning to facilitate the learning process.
By synthesizing the strengths and limitations of these approaches, we aim to provide a roadmap for future research in this exciting and rapidly evolving field.

\subsubsection*{Acknowledgements.}
Rajarshi Roy acknowledges partial funding by the ERC under the European
Union’s Horizon 2020 research and innovation programme
(grant agreement No.834115, FUN2MODEL) and Daniel Neider acknowledges funding by Deutsche Forschungsgemeinschaft (DFG) (grant number 459419731).

%% file: survey.bbl
\begin{thebibliography}{10}

\bibitem{DBLP:conf/fmcad/AlurBJMRSSSTU13}
Rajeev Alur, Rastislav Bod{\'{\i}}k, Garvit Juniwal, Milo M.~K. Martin, Mukund Raghothaman, Sanjit~A. Seshia, Rishabh Singh, Armando Solar{-}Lezama, Emina Torlak, and Abhishek Udupa.
\newblock Syntax-guided synthesis.
\newblock In {\em Formal Methods in Computer-Aided Design, {FMCAD} 2013, Portland, OR, USA, October 20-23, 2013}, pages 1--8. {IEEE}, 2013.
\newblock URL: \url{https://ieeexplore.ieee.org/document/6679385/}.

\bibitem{AmmonsBL02}
Glenn Ammons, Rastislav Bod{\'{\i}}k, and James~R. Larus.
\newblock Mining specifications.
\newblock In John Launchbury and John~C. Mitchell, editors, {\em Conference Record of {POPL} 2002: The 29th {SIGPLAN-SIGACT} Symposium on Principles of Programming Languages, Portland, OR, USA, January 16-18, 2002}, pages 4--16. {ACM}, 2002.
\newblock \href {https://doi.org/10.1145/503272.503275} {\path{doi:10.1145/503272.503275}}.

\bibitem{ArifLERCT20}
M.~Fareed Arif, Daniel Larraz, Mitziu Echeverria, Andrew Reynolds, Omar Chowdhury, and Cesare Tinelli.
\newblock {SYSLITE:} syntax-guided synthesis of {PLTL} formulas from finite traces.
\newblock In {\em {FMCAD}}, pages 93--103. {IEEE}, 2020.

\bibitem{asarin}
Eugene Asarin, Alexandre Donz\'{e}, Oded Maler, and Dejan Nickovic.
\newblock Parametric identification of temporal properties.
\newblock In {\em Proceedings of the Second International Conference on Runtime Verification}, RV'11, page 147–160, Berlin, Heidelberg, 2011. Springer-Verlag.
\newblock \href {https://doi.org/10.1007/978-3-642-29860-8\_12} {\path{doi:10.1007/978-3-642-29860-8\_12}}.

\bibitem{BacheriniFTZ06}
Stefano Bacherini, Alessandro Fantechi, Matteo Tempestini, and Niccol{\`{o}} Zingoni.
\newblock A story about formal methods adoption by a railway signaling manufacturer.
\newblock In {\em {FM}}, volume 4085 of {\em Lecture Notes in Computer Science}, pages 179--189. Springer, 2006.

\bibitem{BadeauA05}
Fr{\'{e}}d{\'{e}}ric Badeau and Arnaud Amelot.
\newblock Using {B} as a high level programming language in an industrial project: Roissy {VAL}.
\newblock In {\em {ZB}}, volume 3455 of {\em Lecture Notes in Computer Science}, pages 334--354. Springer, 2005.

\bibitem{model-checking-book}
Christel Baier and Joost{-}Pieter Katoen.
\newblock {\em Principles of model checking}.
\newblock {MIT} Press, 2008.

\bibitem{DBLP:conf/cav/BarrettCDHJKRT11}
Clark~W. Barrett, Christopher~L. Conway, Morgan Deters, Liana Hadarean, Dejan Jovanovic, Tim King, Andrew Reynolds, and Cesare Tinelli.
\newblock {CVC4}.
\newblock In Ganesh Gopalakrishnan and Shaz Qadeer, editors, {\em Computer Aided Verification - 23rd International Conference, {CAV} 2011, Snowbird, UT, USA, July 14-20, 2011. Proceedings}, volume 6806 of {\em Lecture Notes in Computer Science}, pages 171--177. Springer, 2011.
\newblock \href {https://doi.org/10.1007/978-3-642-22110-1\_14} {\path{doi:10.1007/978-3-642-22110-1\_14}}.

\bibitem{BartocciMNN22}
Ezio Bartocci, Cristinel Mateis, Eleonora Nesterini, and Dejan Nickovic.
\newblock Survey on mining signal temporal logic specifications.
\newblock {\em Inf. Comput.}, 289(Part):104957, 2022.
\newblock URL: \url{https://doi.org/10.1016/j.ic.2022.104957}, \href {https://doi.org/10.1016/J.IC.2022.104957} {\path{doi:10.1016/J.IC.2022.104957}}.

\bibitem{DBLP:journals/jsat/Biere08}
Armin Biere.
\newblock Picosat essentials.
\newblock {\em J. Satisf. Boolean Model. Comput.}, 4(2-4):75--97, 2008.
\newblock URL: \url{https://doi.org/10.3233/sat190039}, \href {https://doi.org/10.3233/SAT190039} {\path{doi:10.3233/SAT190039}}.

\bibitem{BjornerH14}
Dines Bj{\o}rner and Klaus Havelund.
\newblock 40 years of formal methods - some obstacles and some possibilities?
\newblock In {\em {FM}}, volume 8442 of {\em Lecture Notes in Computer Science}, pages 42--61. Springer, 2014.

\bibitem{DBLP:conf/sycss/BjornerP14}
Nikolaj~S. Bj{\o}rner and Anh{-}Dung Phan.
\newblock {\(\nu\)}z - maximal satisfaction with {Z3}.
\newblock In Temur Kutsia and Andrei Voronkov, editors, {\em 6th International Symposium on Symbolic Computation in Software Science, {SCSS} 2014, Gammarth, La Marsa, Tunisia, December 7-8, 2014}, volume~30 of {\em EPiC Series in Computing}, pages 1--9. EasyChair, 2014.
\newblock URL: \url{https://doi.org/10.29007/jmxj}, \href {https://doi.org/10.29007/JMXJ} {\path{doi:10.29007/JMXJ}}.

\bibitem{dtbombara}
Giuseppe Bombara, Cristian-Ioan Vasile, Francisco Penedo, Hirotoshi Yasuoka, and Calin Belta.
\newblock A decision tree approach to data classification using signal temporal logic.
\newblock In {\em Proceedings of the 19th International Conference on Hybrid Systems: Computation and Control}, HSCC '16, page 1–10, New York, NY, USA, 2016. Association for Computing Machinery.
\newblock \href {https://doi.org/10.1145/2883817.2883843} {\path{doi:10.1145/2883817.2883843}}.

\bibitem{dtmethod}
Giuseppe Bombara, Cristian~Ioan Vasile, Francisco Penedo, Hirotoshi Yasuoka, and Calin Belta.
\newblock A decision tree approach to data classification using signal temporal logic.
\newblock In {\em Proceedings of the 19th International Conference on Hybrid Systems: Computation and Control}, HSCC '16, page 1–10, New York, NY, USA, 2016. Association for Computing Machinery.
\newblock \href {https://doi.org/10.1145/2883817.2883843} {\path{doi:10.1145/2883817.2883843}}.

\bibitem{DBLP:journals/corr/abs-2408-04486}
Benjamin Bordais, Daniel Neider, and Rajarshi Roy.
\newblock The complexity of learning temporal properties.
\newblock {\em CoRR}, abs/2408.04486, 2024.
\newblock URL: \url{https://doi.org/10.48550/arXiv.2408.04486}, \href {https://arxiv.org/abs/2408.04486} {\path{arXiv:2408.04486}}, \href {https://doi.org/10.48550/ARXIV.2408.04486} {\path{doi:10.48550/ARXIV.2408.04486}}.

\bibitem{DBLP:conf/fm/BordaisNR24}
Benjamin Bordais, Daniel Neider, and Rajarshi Roy.
\newblock Learning branching-time properties in {CTL} and {ATL} via constraint solving.
\newblock In Andr{\'{e}} Platzer, Kristin~Yvonne Rozier, Matteo Pradella, and Matteo Rossi, editors, {\em Formal Methods - 26th International Symposium, {FM} 2024, Milan, Italy, September 9-13, 2024, Proceedings, Part {I}}, volume 14933 of {\em Lecture Notes in Computer Science}, pages 304--323. Springer, 2024.
\newblock \href {https://doi.org/10.1007/978-3-031-71162-6\_16} {\path{doi:10.1007/978-3-031-71162-6\_16}}.

\bibitem{DBLP:conf/icra/Bozkurt0ZP20}
Alper~Kamil Bozkurt, Yu~Wang, Michael~M. Zavlanos, and Miroslav Pajic.
\newblock Control synthesis from linear temporal logic specifications using model-free reinforcement learning.
\newblock In {\em 2020 {IEEE} International Conference on Robotics and Automation, {ICRA} 2020, Paris, France, May 31 - August 31, 2020}, pages 10349--10355. {IEEE}, 2020.
\newblock \href {https://doi.org/10.1109/ICRA40945.2020.9196796} {\path{doi:10.1109/ICRA40945.2020.9196796}}.

\bibitem{DBLP:conf/ijcai/CamachoIKVM19}
Alberto Camacho, Rodrigo~Toro Icarte, Toryn~Q. Klassen, Richard~Anthony Valenzano, and Sheila~A. McIlraith.
\newblock {LTL} and beyond: Formal languages for reward function specification in reinforcement learning.
\newblock In Sarit Kraus, editor, {\em Proceedings of the Twenty-Eighth International Joint Conference on Artificial Intelligence, {IJCAI} 2019, Macao, China, August 10-16, 2019}, pages 6065--6073. ijcai.org, 2019.
\newblock URL: \url{https://doi.org/10.24963/ijcai.2019/840}, \href {https://doi.org/10.24963/IJCAI.2019/840} {\path{doi:10.24963/IJCAI.2019/840}}.

\bibitem{CamachoM19}
Alberto Camacho and Sheila~A. McIlraith.
\newblock Learning interpretable models expressed in linear temporal logic.
\newblock In {\em {ICAPS}}, pages 621--630. {AAAI} Press, 2019.

\bibitem{Chan00}
William Chan.
\newblock Temporal-logic queries.
\newblock In {\em {CAV}}, volume 1855 of {\em Lecture Notes in Computer Science}, pages 450--463. Springer, 2000.

\bibitem{Cherukuri0S22}
Himaja Cherukuri, Alessio Ferrari, and Paola Spoletini.
\newblock Towards explainable formal methods: From {LTL} to natural language with neural machine translation.
\newblock In Vincenzo Gervasi and Andreas Vogelsang, editors, {\em Requirements Engineering: Foundation for Software Quality - 28th International Working Conference, {REFSQ} 2022, Birmingham, UK, March 21-24, 2022, Proceedings}, volume 13216 of {\em Lecture Notes in Computer Science}, pages 79--86. Springer, 2022.
\newblock \href {https://doi.org/10.1007/978-3-030-98464-9\_7} {\path{doi:10.1007/978-3-030-98464-9\_7}}.

\bibitem{ChouOB22}
Glen Chou, Necmiye Ozay, and Dmitry Berenson.
\newblock Learning temporal logic formulas from suboptimal demonstrations: theory and experiments.
\newblock {\em Auton. Robots}, 46(1):149--174, 2022.
\newblock URL: \url{https://doi.org/10.1007/s10514-021-10004-x}, \href {https://doi.org/10.1007/S10514-021-10004-X} {\path{doi:10.1007/S10514-021-10004-X}}.

\bibitem{DBLP:journals/fmsd/ClarkeBRZ01}
Edmund~M. Clarke, Armin Biere, Richard Raimi, and Yunshan Zhu.
\newblock Bounded model checking using satisfiability solving.
\newblock {\em Formal Methods Syst. Des.}, 19(1):7--34, 2001.
\newblock \href {https://doi.org/10.1023/A:1011276507260} {\path{doi:10.1023/A:1011276507260}}.

\bibitem{CoferM14}
Darren~D. Cofer and Steven~P. Miller.
\newblock {DO-333} certification case studies.
\newblock In {\em {NASA} Formal Methods}, volume 8430 of {\em Lecture Notes in Computer Science}, pages 1--15. Springer, 2014.

\bibitem{CoslerHMST23}
Matthias Cosler, Christopher Hahn, Daniel Mendoza, Frederik Schmitt, and Caroline Trippel.
\newblock nl2spec: Interactively translating unstructured natural language to temporal logics with large language models.
\newblock In Constantin Enea and Akash Lal, editors, {\em Computer Aided Verification - 35th International Conference, {CAV} 2023, Paris, France, July 17-22, 2023, Proceedings, Part {II}}, volume 13965 of {\em Lecture Notes in Computer Science}, pages 383--396. Springer, 2023.
\newblock \href {https://doi.org/10.1007/978-3-031-37703-7\_18} {\path{doi:10.1007/978-3-031-37703-7\_18}}.

\bibitem{DBLP:conf/tacas/MouraB08}
Leonardo~Mendon{\c{c}}a de~Moura and Nikolaj~S. Bj{\o}rner.
\newblock {Z3:} an efficient {SMT} solver.
\newblock In C.~R. Ramakrishnan and Jakob Rehof, editors, {\em Tools and Algorithms for the Construction and Analysis of Systems, 14th International Conference, {TACAS} 2008, Held as Part of the Joint European Conferences on Theory and Practice of Software, {ETAPS} 2008, Budapest, Hungary, March 29-April 6, 2008. Proceedings}, volume 4963 of {\em Lecture Notes in Computer Science}, pages 337--340. Springer, 2008.
\newblock \href {https://doi.org/10.1007/978-3-540-78800-3\_24} {\path{doi:10.1007/978-3-540-78800-3\_24}}.

\bibitem{DBLP:conf/fmsp/DwyerAC98}
Matthew~B. Dwyer, George~S. Avrunin, and James~C. Corbett.
\newblock Property specification patterns for finite-state verification.
\newblock In Mark~A. Ardis and Joanne~M. Atlee, editors, {\em Proceedings of the Second Workshop on Formal Methods in Software Practice, March 4-5, 1998, Clearwater Beach, Florida, {USA}}, pages 7--15. {ACM}, 1998.
\newblock \href {https://doi.org/10.1145/298595.298598} {\path{doi:10.1145/298595.298598}}.

\bibitem{FeckoUASDMM00}
Mariusz~A. Fecko, M.~{\"{U}}mit Uyar, Paul~D. Amer, Adarshpal~S. Sethi, Theodore Dzik, R.~Menell, and Michael McMahon.
\newblock A success story of formal description techniques: Estelle specification and test generation for {MIL-STD} 188-220.
\newblock {\em Comput. Commun.}, 23(12):1196--1213, 2000.

\bibitem{DBLP:conf/icgi/FijalkowL21}
Nathana{\"{e}}l Fijalkow and Guillaume Lagarde.
\newblock The complexity of learning linear temporal formulas from examples.
\newblock In {\em {ICGI}}, volume 153 of {\em Proceedings of Machine Learning Research}, pages 237--250. {PMLR}, 2021.

\bibitem{FinucaneJK10}
Cameron Finucane, Gangyuan Jing, and Hadas Kress{-}Gazit.
\newblock Ltlmop: Experimenting with language, temporal logic and robot control.
\newblock In {\em 2010 {IEEE/RSJ} International Conference on Intelligent Robots and Systems, October 18-22, 2010, Taipei, Taiwan}, pages 1988--1993. {IEEE}, 2010.
\newblock \href {https://doi.org/10.1109/IROS.2010.5650371} {\path{doi:10.1109/IROS.2010.5650371}}.

\bibitem{nl2ltl}
Francesco Fuggitti and Tathagata Chakraborti.
\newblock Nl2ltl - a python package for converting natural language (nl) instructions to linear temporal logic (ltl) formulas.
\newblock In {\em Proceedings of the Thirty-Seventh AAAI Conference on Artificial Intelligence and Thirty-Fifth Conference on Innovative Applications of Artificial Intelligence and Thirteenth Symposium on Educational Advances in Artificial Intelligence}, AAAI'23/IAAI'23/EAAI'23. AAAI Press, 2023.
\newblock \href {https://doi.org/10.1609/aaai.v37i13.27068} {\path{doi:10.1609/aaai.v37i13.27068}}.

\bibitem{GaglioneNRTX21}
Jean{-}Rapha{\"{e}}l Gaglione, Daniel Neider, Rajarshi Roy, Ufuk Topcu, and Zhe Xu.
\newblock Learning linear temporal properties from noisy data: {A} maxsat-based approach.
\newblock In Zhe Hou and Vijay Ganesh, editors, {\em Automated Technology for Verification and Analysis - 19th International Symposium, {ATVA} 2021, Gold Coast, QLD, Australia, October 18-22, 2021, Proceedings}, volume 12971 of {\em Lecture Notes in Computer Science}, pages 74--90. Springer, 2021.
\newblock \href {https://doi.org/10.1007/978-3-030-88885-5\_6} {\path{doi:10.1007/978-3-030-88885-5\_6}}.

\bibitem{GarioCMTR16}
Marco Gario, Alessandro Cimatti, Cristian Mattarei, Stefano Tonetta, and Kristin~Yvonne Rozier.
\newblock Model checking at scale: Automated air traffic control design space exploration.
\newblock In {\em {CAV} {(2)}}, volume 9780 of {\em Lecture Notes in Computer Science}, pages 3--22. Springer, 2016.

\bibitem{DBLP:series/synthesis/2012Gebser}
Martin Gebser, Roland Kaminski, Benjamin Kaufmann, and Torsten Schaub.
\newblock {\em Answer Set Solving in Practice}.
\newblock Synthesis Lectures on Artificial Intelligence and Machine Learning. Morgan {\&} Claypool Publishers, 2012.
\newblock \href {https://doi.org/10.2200/S00457ED1V01Y201211AIM019} {\path{doi:10.2200/S00457ED1V01Y201211AIM019}}.

\bibitem{DBLP:journals/ngc/GelfondL91}
Michael Gelfond and Vladimir Lifschitz.
\newblock Classical negation in logic programs and disjunctive databases.
\newblock {\em New Gener. Comput.}, 9(3/4):365--386, 1991.
\newblock \href {https://doi.org/10.1007/BF03037169} {\path{doi:10.1007/BF03037169}}.

\bibitem{GhiorziCPBTN23}
Enrico Ghiorzi, Michele Colledanchise, Gianluca Piquet, Stefano Bernagozzi, Armando Tacchella, and Lorenzo Natale.
\newblock Learning linear temporal properties for autonomous robotic systems.
\newblock {\em {IEEE} Robotics Autom. Lett.}, 8(5):2930--2937, 2023.
\newblock \href {https://doi.org/10.1109/LRA.2023.3263368} {\path{doi:10.1109/LRA.2023.3263368}}.

\bibitem{GhoshELLSS16}
Shalini Ghosh, Daniel Elenius, Wenchao Li, Patrick Lincoln, Natarajan Shankar, and Wilfried Steiner.
\newblock {ARSENAL:} automatic requirements specification extraction from natural language.
\newblock In {\em {NASA} Formal Methods, NFM}, 2016.
\newblock \href {https://doi.org/10.1007/978-3-319-40648-0\_4} {\path{doi:10.1007/978-3-319-40648-0\_4}}.

\bibitem{GiacomoV13}
Giuseppe~De Giacomo and Moshe~Y. Vardi.
\newblock Linear temporal logic and linear dynamic logic on finite traces.
\newblock In Francesca Rossi, editor, {\em {IJCAI} 2013, Proceedings of the 23rd International Joint Conference on Artificial Intelligence, Beijing, China, August 3-9, 2013}, pages 854--860. {IJCAI/AAAI}, 2013.
\newblock URL: \url{http://www.aaai.org/ocs/index.php/IJCAI/IJCAI13/paper/view/6997}.

\bibitem{GiannakopoulouP20a}
Dimitra Giannakopoulou, Thomas Pressburger, Anastasia Mavridou, Julian Rhein, Johann Schumann, and Nija Shi.
\newblock Formal requirements elicitation with {FRET}.
\newblock In {\em International Conference on Requirements Engineering: Foundation for Software Quality, REFSQ}, 2020.
\newblock URL: \url{http://ceur-ws.org/Vol-2584/PT-paper4.pdf}.

\bibitem{GreenmanSNK23}
Ben Greenman, Sam Saarinen, Tim Nelson, and Shriram Krishnamurthi.
\newblock Little tricky logic: Misconceptions in the understanding of {LTL}.
\newblock {\em Art Sci. Eng. Program.}, 7(2), 2023.

\bibitem{HahnSKRF21}
Christopher Hahn, Frederik Schmitt, Jens~U. Kreber, Markus~Norman Rabe, and Bernd Finkbeiner.
\newblock Teaching temporal logics to neural networks.
\newblock In {\em 9th International Conference on Learning Representations, {ICLR} 2021, Virtual Event, Austria, May 3-7, 2021}. OpenReview.net, 2021.
\newblock URL: \url{https://openreview.net/forum?id=dOcQK-f4byz}.

\bibitem{DBLP:journals/debu/HamiltonYL17}
William~L. Hamilton, Rex Ying, and Jure Leskovec.
\newblock Representation learning on graphs: Methods and applications.
\newblock {\em {IEEE} Data Eng. Bull.}, 40(3):52--74, 2017.
\newblock URL: \url{http://sites.computer.org/debull/A17sept/p52.pdf}.

\bibitem{DBLP:conf/cdc/HasanbeigKAKPL19}
Mohammadhosein Hasanbeig, Yiannis Kantaros, Alessandro Abate, Daniel Kroening, George~J. Pappas, and Insup Lee.
\newblock Reinforcement learning for temporal logic control synthesis with probabilistic satisfaction guarantees.
\newblock In {\em 58th {IEEE} Conference on Decision and Control, {CDC} 2019, Nice, France, December 11-13, 2019}, pages 5338--5343. {IEEE}, 2019.
\newblock \href {https://doi.org/10.1109/CDC40024.2019.9028919} {\path{doi:10.1109/CDC40024.2019.9028919}}.

\bibitem{Holzmann02}
Gerard~J. Holzmann.
\newblock The logic of bugs.
\newblock In {\em {SIGSOFT} {FSE}}, pages 81--87. {ACM}, 2002.

\bibitem{IeloLFRGR23}
Antonio Ielo, Mark Law, Valeria Fionda, Francesco Ricca, Giuseppe De~Giacomo, and Alessandra Russo.
\newblock Towards ilp-based ltlf passive learning.
\newblock In {\em Inductive Logic Programming: 32nd International Conference, ILP 2023, Bari, Italy, November 13–15, 2023, Proceedings}, page 30–45, Berlin, Heidelberg, 2023. Springer-Verlag.
\newblock \href {https://doi.org/10.1007/978-3-031-49299-0_3} {\path{doi:10.1007/978-3-031-49299-0_3}}.

\bibitem{KimMSAS19}
Joseph Kim, Christian Muise, Ankit Shah, Shubham Agarwal, and Julie Shah.
\newblock Bayesian inference of linear temporal logic specifications for contrastive explanations.
\newblock In {\em {IJCAI}}, pages 5591--5598. ijcai.org, 2019.

\bibitem{KleinAEHCDEEKNSTW10}
Gerwin Klein, June Andronick, Kevin Elphinstone, Gernot Heiser, David Cock, Philip Derrin, Dhammika Elkaduwe, Kai Engelhardt, Rafal Kolanski, Michael Norrish, Thomas Sewell, Harvey Tuch, and Simon Winwood.
\newblock sel4: formal verification of an operating-system kernel.
\newblock {\em Commun. {ACM}}, 53(6):107--115, 2010.

\bibitem{rpstl2}
Zhaodan Kong, Austin Jones, and Calin Belta.
\newblock Temporal logics for learning and detection of anomalous behavior.
\newblock {\em IEEE Transactions on Automatic Control}, 62(3):1210--1222, 2017.
\newblock \href {https://doi.org/10.1109/TAC.2016.2585083} {\path{doi:10.1109/TAC.2016.2585083}}.

\bibitem{rpstl1}
Zhaodan Kong, Austin Jones, Ana Medina~Ayala, Ebru Aydin~Gol, and Calin Belta.
\newblock Temporal logic inference for classification and prediction from data.
\newblock In {\em Proceedings of the 17th International Conference on Hybrid Systems: Computation and Control}, HSCC '14, page 273–282, New York, NY, USA, 2014. Association for Computing Machinery.
\newblock \href {https://doi.org/10.1145/2562059.2562146} {\path{doi:10.1145/2562059.2562146}}.

\bibitem{Kress-GazitFP08}
Hadas Kress{-}Gazit, Georgios~E. Fainekos, and George~J. Pappas.
\newblock Translating structured english to robot controllers.
\newblock {\em Advanced Robotics}, 22(12):1343--1359, 2008.

\bibitem{DBLP:conf/kbse/LemieuxB15}
Caroline Lemieux and Ivan Beschastnikh.
\newblock Investigating program behavior using the texada {LTL} specifications miner.
\newblock In Myra~B. Cohen, Lars Grunske, and Michael Whalen, editors, {\em 30th {IEEE/ACM} International Conference on Automated Software Engineering, {ASE} 2015, Lincoln, NE, USA, November 9-13, 2015}, pages 870--875. {IEEE} Computer Society, 2015.
\newblock \href {https://doi.org/10.1109/ASE.2015.94} {\path{doi:10.1109/ASE.2015.94}}.

\bibitem{LiDS11}
Wenchao Li, Lili Dworkin, and Sanjit~A. Seshia.
\newblock Mining assumptions for synthesis.
\newblock In {\em {MEMOCODE}}, pages 43--50. {IEEE}, 2011.

\bibitem{DBLP:conf/iros/LiVB17}
Xiao Li, Cristian~Ioan Vasile, and Calin Belta.
\newblock Reinforcement learning with temporal logic rewards.
\newblock In {\em 2017 {IEEE/RSJ} International Conference on Intelligent Robots and Systems, {IROS} 2017, Vancouver, BC, Canada, September 24-28, 2017}, pages 3834--3839. {IEEE}, 2017.
\newblock \href {https://doi.org/10.1109/IROS.2017.8206234} {\path{doi:10.1109/IROS.2017.8206234}}.

\bibitem{lang2LTL}
Jason~Xinyu Liu, Ziyi Yang, Ifrah Idrees, Sam Liang, Benjamin Schornstein, Stefanie Tellex, and Ankit Shah.
\newblock Lang2ltl: Translating natural language commands to temporal robot task specification.
\newblock {\em CoRR}, abs/2302.11649, 2023.
\newblock URL: \url{https://doi.org/10.48550/arXiv.2302.11649}, \href {https://arxiv.org/abs/2302.11649} {\path{arXiv:2302.11649}}, \href {https://doi.org/10.48550/ARXIV.2302.11649} {\path{doi:10.48550/ARXIV.2302.11649}}.

\bibitem{Lowe96}
Gavin Lowe.
\newblock Breaking and fixing the needham-schroeder public-key protocol using {FDR}.
\newblock {\em Softw. Concepts Tools}, 17(3):93--102, 1996.

\bibitem{DBLP:conf/aaai/LuoLDWPZ22}
Weilin Luo, Pingjia Liang, Jianfeng Du, Hai Wan, Bo~Peng, and Delong Zhang.
\newblock Bridging ltlf inference to {GNN} inference for learning ltlf formulae.
\newblock In {\em {AAAI}}, pages 9849--9857. {AAAI} Press, 2022.

\bibitem{LutzNR23}
Simon Lutz, Daniel Neider, and Rajarshi Roy.
\newblock Specification sketching for linear temporal logic.
\newblock In {\'{E}}tienne Andr{\'{e}} and Jun Sun, editors, {\em Automated Technology for Verification and Analysis - 21st International Symposium, {ATVA} 2023, Singapore, October 24-27, 2023, Proceedings, Part {II}}, volume 14216 of {\em Lecture Notes in Computer Science}, pages 26--48. Springer, 2023.
\newblock \href {https://doi.org/10.1007/978-3-031-45332-8\_2} {\path{doi:10.1007/978-3-031-45332-8\_2}}.

\bibitem{DBLP:journals/corr/abs-2312-16336}
Corto Mascle, Nathana{\"{e}}l Fijalkow, and Guillaume Lagarde.
\newblock Learning temporal formulas from examples is hard.
\newblock {\em CoRR}, abs/2312.16336, 2023.
\newblock URL: \url{https://doi.org/10.48550/arXiv.2312.16336}, \href {https://arxiv.org/abs/2312.16336} {\path{arXiv:2312.16336}}, \href {https://doi.org/10.48550/ARXIV.2312.16336} {\path{doi:10.48550/ARXIV.2312.16336}}.

\bibitem{MohammadinejadD20}
Sara Mohammadinejad, Jyotirmoy~V. Deshmukh, Aniruddh~Gopinath Puranic, Marcell Vazquez{-}Chanlatte, and Alexandre Donz{\'{e}}.
\newblock Interpretable classification of time-series data using efficient enumerative techniques.
\newblock In {\em {HSCC} '20: 23rd {ACM} International Conference on Hybrid Systems: Computation and Control, Sydney, New South Wales, Australia, April 21-24, 2020}, pages 9:1--9:10. {ACM}, 2020.
\newblock \href {https://doi.org/10.1145/3365365.3382218} {\path{doi:10.1145/3365365.3382218}}.

\bibitem{flie}
Daniel Neider and Ivan Gavran.
\newblock Learning linear temporal properties.
\newblock In Nikolaj~S. Bj{\o}rner and Arie Gurfinkel, editors, {\em 2018 Formal Methods in Computer Aided Design, {FMCAD} 2018, Austin, TX, USA, October 30 - November 2, 2018}, pages 1--10. {IEEE}, 2018.
\newblock \href {https://doi.org/10.23919/FMCAD.2018.8603016} {\path{doi:10.23919/FMCAD.2018.8603016}}.

\bibitem{jpk2025}
Daniel Neider and Rajarshi Roy.
\newblock {\em What Is Formal Verification Without Specifications? A Survey on Mining LTL Specifications}, pages 109--125.
\newblock Springer Nature Switzerland, Cham, 2025.
\newblock \href {https://doi.org/10.1007/978-3-031-75778-5_6} {\path{doi:10.1007/978-3-031-75778-5_6}}.

\bibitem{DBLP:books/daglib/0090563}
George~L. Nemhauser and Laurence~A. Wolsey.
\newblock {\em Integer and Combinatorial Optimization}.
\newblock Wiley interscience series in discrete mathematics and optimization. Wiley, 1988.
\newblock \href {https://doi.org/10.1002/9781118627372} {\path{doi:10.1002/9781118627372}}.

\bibitem{genetic}
Laura Nenzi, Simone Silvetti, Ezio Bartocci, and Luca Bortolussi.
\newblock A robust genetic algorithm for learning temporal specifications from data.
\newblock In Annabelle McIver and Andras Horvath, editors, {\em Quantitative Evaluation of Systems}, pages 323--338, Cham, 2018. Springer International Publishing.

\bibitem{NikoraB09}
Allen~P. Nikora and Galen Balcom.
\newblock Automated identification of {LTL} patterns in natural language requirements.
\newblock In {\em {ISSRE} 2009, 20th International Symposium on Software Reliability Engineering, Mysuru, Karnataka, India, 16-19 November 2009}, pages 185--194. {IEEE} Computer Society, 2009.
\newblock \href {https://doi.org/10.1109/ISSRE.2009.15} {\path{doi:10.1109/ISSRE.2009.15}}.

\bibitem{OhPNHPT19}
Yoonseon Oh, Roma Patel, Thao Nguyen, Baichuan Huang, Ellie Pavlick, and Stefanie Tellex.
\newblock Planning with state abstractions for non-markovian task specifications.
\newblock In Antonio Bicchi, Hadas Kress{-}Gazit, and Seth Hutchinson, editors, {\em Robotics: Science and Systems XV, University of Freiburg, Freiburg im Breisgau, Germany, June 22-26, 2019}, 2019.
\newblock \href {https://doi.org/10.15607/RSS.2019.XV.059} {\path{doi:10.15607/RSS.2019.XV.059}}.

\bibitem{PanCB23}
Jiayi Pan, Glen Chou, and Dmitry Berenson.
\newblock Data-efficient learning of natural language to linear temporal logic translators for robot task specification.
\newblock In {\em {IEEE} International Conference on Robotics and Automation, {ICRA} 2023, London, UK, May 29 - June 2, 2023}, pages 11554--11561. {IEEE}, 2023.
\newblock \href {https://doi.org/10.1109/ICRA48891.2023.10161125} {\path{doi:10.1109/ICRA48891.2023.10161125}}.

\bibitem{DBLP:conf/focs/Pnueli77}
Amir Pnueli.
\newblock The temporal logic of programs.
\newblock In {\em 18th Annual Symposium on Foundations of Computer Science, Providence, Rhode Island, USA, 31 October - 1 November 1977}, pages 46--57. {IEEE} Computer Society, 1977.
\newblock \href {https://doi.org/10.1109/SFCS.1977.32} {\path{doi:10.1109/SFCS.1977.32}}.

\bibitem{DBLP:conf/ijcar/PommelletSS24}
Adrien Pommellet, Daniel Stan, and Simon Scatton.
\newblock Sat-based learning of computation tree logic.
\newblock In Christoph Benzm{\"{u}}ller, Marijn J.~H. Heule, and Renate~A. Schmidt, editors, {\em Automated Reasoning - 12th International Joint Conference, {IJCAR} 2024, Nancy, France, July 3-6, 2024, Proceedings, Part {I}}, volume 14739 of {\em Lecture Notes in Computer Science}, pages 366--385. Springer, 2024.
\newblock \href {https://doi.org/10.1007/978-3-031-63498-7\_22} {\path{doi:10.1007/978-3-031-63498-7\_22}}.

\bibitem{scarlet}
Ritam Raha, Rajarshi Roy, Nathana{\"e}l Fijalkow, and Daniel Neider.
\newblock Scalable anytime algorithms for learning fragments of linear temporal logic.
\newblock In Dana Fisman and Grigore Rosu, editors, {\em Tools and Algorithms for the Construction and Analysis of Systems}, pages 263--280, Cham, 2022. Springer International Publishing.

\bibitem{DBLP:journals/jossw/RahaRFN24}
Ritam Raha, Rajarshi Roy, Nathana{\"{e}}l Fijalkow, and Daniel Neider.
\newblock Scarlet: Scalable anytime algorithms for learning fragments of linear temporal logic.
\newblock {\em J. Open Source Softw.}, 9(93):5052, 2024.
\newblock URL: \url{https://doi.org/10.21105/joss.05052}, \href {https://doi.org/10.21105/JOSS.05052} {\path{doi:10.21105/JOSS.05052}}.

\bibitem{DBLP:conf/vmcai/RahaRFNP24}
Ritam Raha, Rajarshi Roy, Nathana{\"{e}}l Fijalkow, Daniel Neider, and Guillermo~A. P{\'{e}}rez.
\newblock Synthesizing efficiently monitorable formulas in metric temporal logic.
\newblock In {\em {VMCAI} {(2)}}, volume 14500 of {\em Lecture Notes in Computer Science}, pages 264--288. Springer, 2024.

\bibitem{DBLP:conf/cav/ReynoldsBNBT19}
Andrew Reynolds, Haniel Barbosa, Andres N{\"{o}}tzli, Clark~W. Barrett, and Cesare Tinelli.
\newblock cvc4sy: Smart and fast term enumeration for syntax-guided synthesis.
\newblock In Isil Dillig and Serdar Tasiran, editors, {\em Computer Aided Verification - 31st International Conference, {CAV} 2019, New York City, NY, USA, July 15-18, 2019, Proceedings, Part {II}}, volume 11562 of {\em Lecture Notes in Computer Science}, pages 74--83. Springer, 2019.
\newblock \href {https://doi.org/10.1007/978-3-030-25543-5\_5} {\path{doi:10.1007/978-3-030-25543-5\_5}}.

\bibitem{Riener19}
Heinz Riener.
\newblock Exact synthesis of {LTL} properties from traces.
\newblock In {\em {FDL}}, pages 1--6. {IEEE}, 2019.

\bibitem{0002FN20}
Rajarshi Roy, Dana Fisman, and Daniel Neider.
\newblock Learning interpretable models in the property specification language.
\newblock In {\em {IJCAI}}, pages 2213--2219. ijcai.org, 2020.

\bibitem{ltl-from-positive-only}
Rajarshi Roy, Jean{-}Rapha{\"{e}}l Gaglione, Nasim Baharisangari, Daniel Neider, Zhe Xu, and Ufuk Topcu.
\newblock Learning interpretable temporal properties from positive examples only.
\newblock {\em CoRR}, abs/2209.02650, 2022.

\bibitem{Rozier16}
Kristin~Yvonne Rozier.
\newblock Specification: The biggest bottleneck in formal methods and autonomy.
\newblock In {\em {VSTTE}}, volume 9971 of {\em Lecture Notes in Computer Science}, pages 8--26, 2016.

\bibitem{ShahKSL18}
Ankit Shah, Pritish Kamath, Julie~A. Shah, and Shen Li.
\newblock Bayesian inference of temporal task specifications from demonstrations.
\newblock In {\em NeurIPS}, pages 3808--3817, 2018.

\bibitem{LTL-GPU}
Mojtaba Valizadeh, Nathana{\"e}l Fijalkow, and Martin Berger.
\newblock Ltl learning on gpus.
\newblock In Arie Gurfinkel and Vijay Ganesh, editors, {\em Computer Aided Verification}, pages 209--231, Cham, 2024. Springer Nature Switzerland.
\newblock URL: \url{https://doi.org/10.1007/978-3-031-65633-0_10}.

\bibitem{VerhulstJ07}
Eric Verhulst and Gjalt~G. de~Jong.
\newblock Opencomrtos: An ultra-small network centric embedded {RTOS} designed using formal modeling.
\newblock In {\em {SDL} Forum}, volume 4745 of {\em Lecture Notes in Computer Science}, pages 258--271. Springer, 2007.

\bibitem{WachterB06}
Andreas W{\"{a}}chter and Lorenz~T. Biegler.
\newblock On the implementation of an interior-point filter line-search algorithm for large-scale nonlinear programming.
\newblock {\em Math. Program.}, 106(1):25--57, 2006.
\newblock URL: \url{https://doi.org/10.1007/s10107-004-0559-y}, \href {https://doi.org/10.1007/S10107-004-0559-Y} {\path{doi:10.1007/S10107-004-0559-Y}}.

\bibitem{DBLP:conf/aaai/WanLDLYP24}
Hai Wan, Pingjia Liang, Jianfeng Du, Weilin Luo, Rongzhen Ye, and Bo~Peng.
\newblock End-to-end learning of ltlf formulae by faithful ltlf encoding.
\newblock In {\em {AAAI}}, pages 9071--9079. {AAAI} Press, 2024.

\bibitem{WasylkowskiZ11}
Andrzej Wasylkowski and Andreas Zeller.
\newblock Mining temporal specifications from object usage.
\newblock {\em Autom. Softw. Eng.}, 18(3-4):263--292, 2011.

\end{thebibliography}
